\documentclass[reprint,
superscriptaddress,
 amsmath,amssymb,
 aps,
 prl,
]{revtex4-2}

\usepackage{graphicx}
\usepackage[english]{babel}
\usepackage{dcolumn}
\usepackage{bm}
\usepackage{xcolor}
\usepackage{comment}
\usepackage{tabularx}
\usepackage[colorlinks=true, allcolors=blue]{hyperref}
\usepackage{url}

\begin{document}

\preprint{APS/123-QED}

\title{High-precision $Q$-value measurement confirms the potential of $^{135}$Cs 
for antineutrino-mass detection}

\author{A.~de~Roubin}
 \email{antoine.a.deroubin@jyu.fi}
\author{J.~Kostensalo}%
\author{T.~Eronen}%
\author{L.~Canete}
\author{R.~P.~de~Groote}
\author{A.~Jokinen}
\author{A.~Kankainen}
\author{D.~A.~Nesterenko}
\author{I.~D.~Moore}
\author{S.~Rinta-Antila}
\author{J.~Suhonen}
\author{M.~Vil\'en}
\affiliation{%
University of Jyv\"askyl\"a, P.O. Box 35, FI-40014 University of Jyv\"askyl\"a, Finland
}

\date{\today}

\begin{abstract}
The ground-state-to-ground-state $\beta$-decay $Q$-value of $^{135}\textrm{Cs}(7/2^+)\to\,^{135}\textrm{Ba}(3/2^+)$ was directly measured for the first time utilizing the Phase-Imaging Ion-Cyclotron Resonance (PI-ICR) technique at the JYFLTRAP Penning-trap setup. It is the first direct determination of this $Q$-value and its value of 268.66(30)\,keV is a factor of three more precise than the currently adopted $Q$-value in the Atomic Mass Evaluation 2016. Moreover, the $Q$-value deduced from the $\beta$-decay endpoint energy has been found to deviate from our result by approximately 6 standard deviations. The measurement confirms that the first-forbidden unique $\beta^-$-decay transition $^{135}\textrm{Cs}(7/2^+)\to\,^{135}\textrm{Ba}(11/2^-)$ is a candidate for antineutrino-mass measurements with an ultra-low $Q$-value of $0.44(31)$ keV. This $Q$-value is almost an order of magnitude smaller than in any presently running or planned direct (anti)neutrino-mass experiment.
\end{abstract}

\maketitle


The determination of the absolute scale of the (anti)neutrino mass is one of the most important and intriguing goals in particle physics. This can be addressed by measurements of neutrinoless double $\beta$-decay via the effective neutrino mass~\cite{Ejiri2019} or by $\beta$-decay experiments via the electron (anti)neutrino mass~\cite{Aker2019}. The latter experiments are model-independent methods to directly measure the mass from the distortion of the $\beta$-electron spectrum end point, like in the KATRIN (KArlsruhe TRitium Neutrino)~\cite{Aker2019} and MARE (Microcalorimeter Arrays for a Rhenium Experiment)~\cite{Ferri2015} experiments or from the total-absorption-spectrum end point, like in the ECHo (Electron Capture in $^{163}$Ho) experiment~\cite{Gastaldo2017}.
In these experiments one strives for sub-eV mass sensitivity, which necessitates the use of nuclear decays of as small as possible decay energy ($Q$-value) in order to reduce the background at the end point, from which the neutrino mass is extracted.
The corresponding $Q$-values are $Q_{\beta}=18.5718(12)$\,keV for KATRIN~\cite{Otten_2008}, $Q_{\beta}=2.4666(16)$\,keV for MARE~\cite{Nesterenko_2014} and $Q_{\rm EC}=2.858(10)(50)$\,keV for ECHo~\cite{Ranitzsch2017} with its statistical and systematic uncertainties, respectively.

\begin{figure}[h!]
 \begin{center}
    \includegraphics[width=\columnwidth]{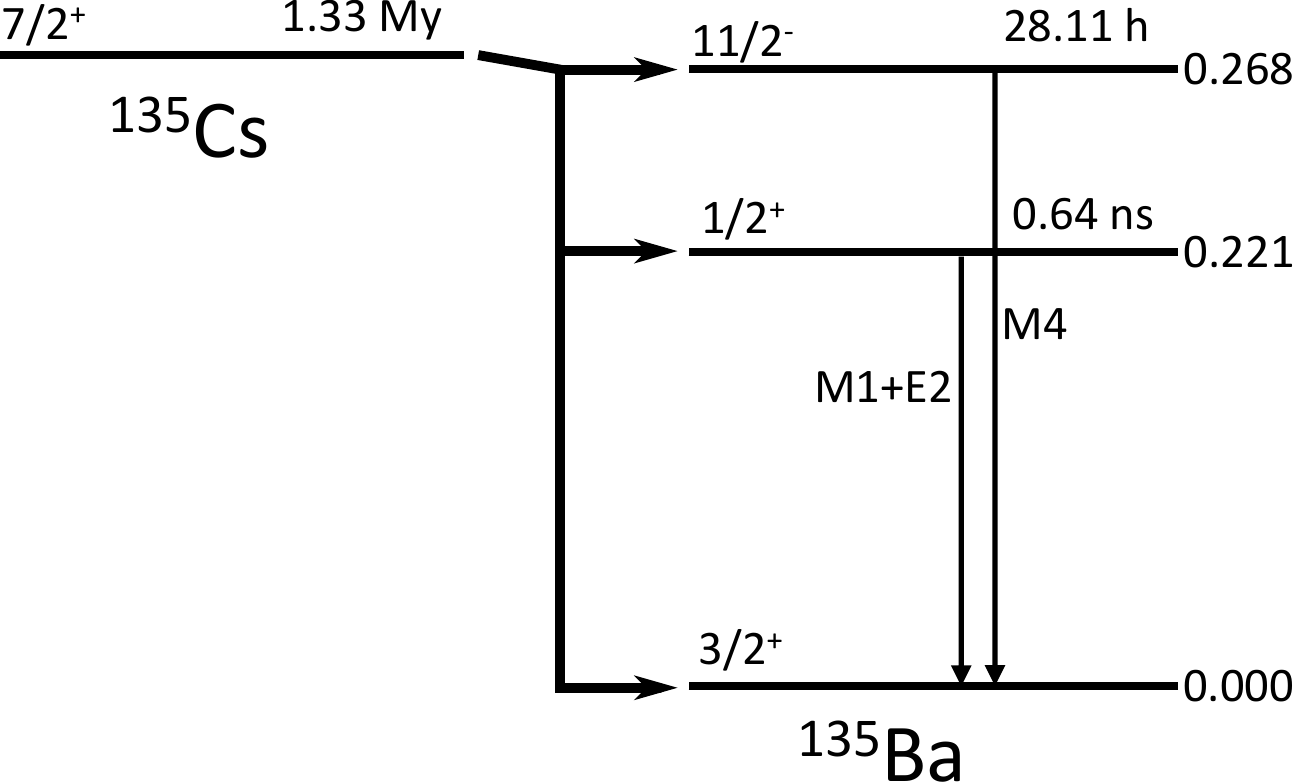}
    \caption{\label{fig:decay} $\beta^-$-decay of the ground state of $^{135}$Cs to the ground state and first two excites states in $^{135}$Ba. The yet undetected transition to the second excited state ($11/2^-$) is an ultra-low $Q$-value transition studied in this work. The transition to the first excited state is greatly hindered by the large change in angular momentum. The numbers to the right of the energy levels are excitation energies in MeV.}
 \end{center}
\end{figure}

The $\beta$-decay of the $7/2^+$ state of $^{135}$Cs to the $11/2^-$ state of $^{135}$Ba has been proposed as a candidate for (anti)neutrino-mass measurements, see ~\cite{Mustonen2011}. However, it has not been clear if it is energetically allowed. The decay has never been observed directly and the low $Q$-value has been deduced from the well known excitation energy of the $11/2^-$ state at 268.218(20)\,keV in $^{135}$Ba (see Fig. ~\ref{fig:decay}) and the ground-state-to-ground-state (GS-to-GS) $\beta$-decay $Q$-value. Two GS-to-GS $Q$-values exist in the literature. One is 210(10)\,keV, which is based on $\beta$-decay endpoint energy~\cite{Sugarman1949}. In case this value is correct, the decay to the $11/2^-$ state of $^{135}$Ba would be energetically forbidden with a $Q$-value of -58(10)\,keV. The other GS-to-GS $Q$-value available in the literature is based from the AME2016~\cite{Wang_2017} and is equal to 268.9(10)\,keV. From this, the decay to the $11/2^-$ state has a $Q$-value of 0.5(12)\,keV. Although the value is more precise, it does not reliably exclude whether the decay to the $11/2^-$~state is energetically possible or not. In~\cite{Mustonen2011} both $Q$-value scenarios were discussed with implications to the partial half-lives of the decays to the excited states in $^{135}$Ba. 
In this Letter, we report on the first direct $Q$-value measurement of the $\beta$-decay of $^{135}$Cs in order to verify 
whether the transition $^{135}\textrm{Cs}(7/2^+)\to\,^{135}\textrm{Ba}(11/2^-)$ could serve as a potential candidate for very low $Q$-value antineutrino-mass measurements.

\begin{figure}[h!]
 \begin{center}
    \includegraphics[width=\linewidth]{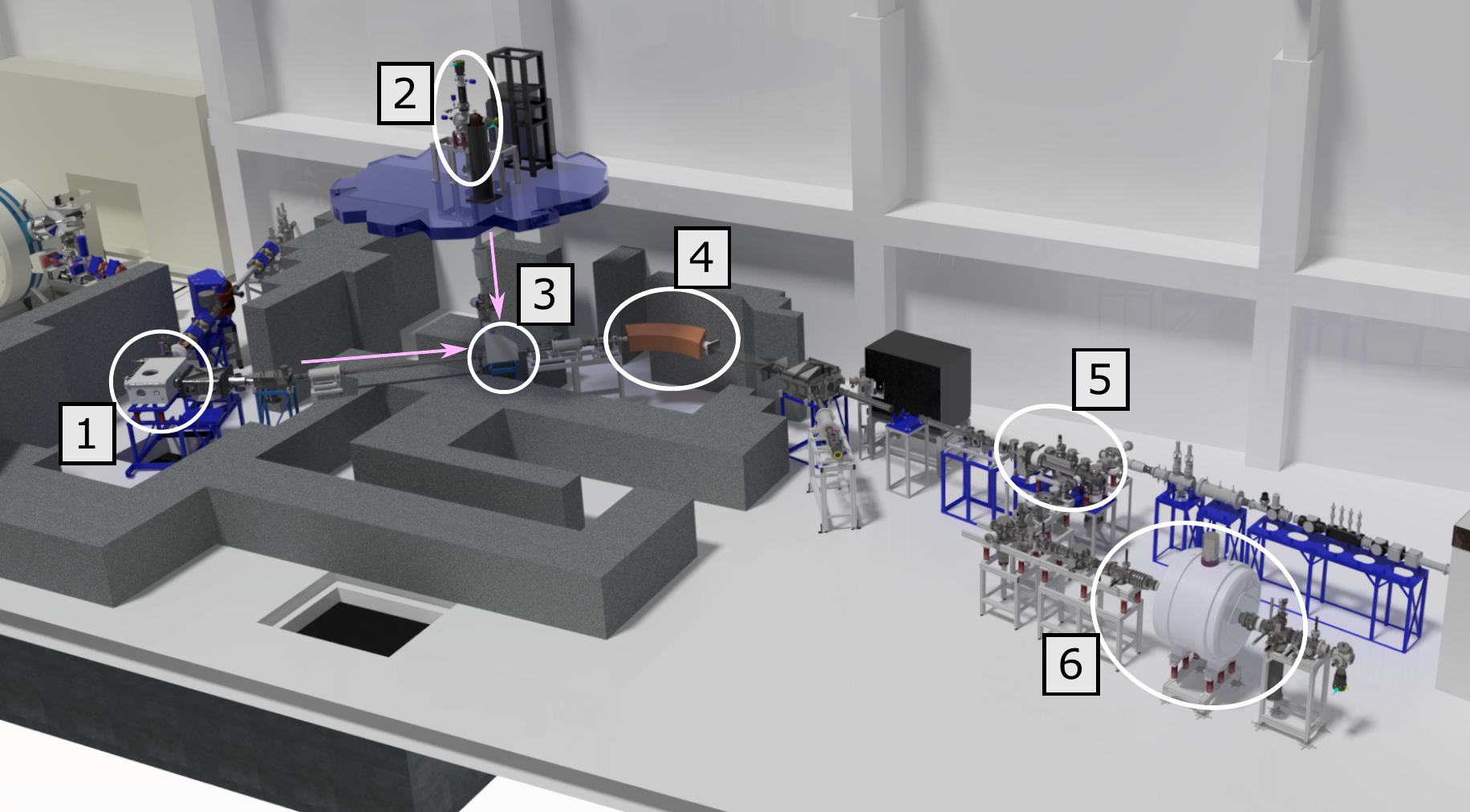}
    \caption{\label{fig:IG_layout}Layout of the IGISOL facility. The radioactive $^{135}$Cs$^+$ ions were produced with proton-induced fission reactions (1), the stable $^{135}$Ba$^+$ ions with an off-line source (2). The beam from either source was selected with an electrostatic kicker (3). The mass number selection was performed with a dipole magnet (4), the ion bunching with the cooler-buncher (5) and finally the mass-difference measurement with the JYFLTRAP Penning trap setup (6). }
 \end{center}
\end{figure}

The GS-to-GS $Q$-value of $^{135}$Cs was measured 
using the JYFLTRAP double Penning trap setup mass spectrometer at the Ion Guide Isotope Separator On-Line (IGISOL) facility~\cite{Aysto2001,Moore2013}, see Fig.~\ref{fig:IG_layout}. The $^{135}$Cs$(7/2^+)$ ions were produced using proton-induced fission with a 50-MeV proton beam impinging into a $^\mathrm{nat}$U target. The reference $^{135}$Ba$(3/2^+)$ ions were separately produced with an off-line glow-discharge ion source~\cite{2020_VILEN}. For high precision Penning trap mass measurements it is of utmost importance to have a mono-isotopic sample of ions. Since it is not possible to separate $^{135}$Ba$(11/2^-)$ and $^{135}$Cs$(7/2^+)$ that have nearly identical mass with currently available separation techniques~\cite{2008_Eronen, Nesterenko2018}, fission reaction was chosen to produce $^{135}$Cs ions. Based on a semi-empirical fit to the independent fission yield data to theoretical models \cite{2016_Pentilla}, the $^{135}$Ba$(11/2^-)$ yield was expected to be a factor of 100 less than $^{135}$Cs$(7/2^+)$.


The ion beams, irrespective of the source, were coarsely mass separated to contain only $A/q = 135$ ions with a dipole magnet, where $q$ is the charge state of the ion, and injected into the radio-frequency cooler-buncher. The resulting bunched beams were delivered to the purification Penning trap, where the ions were selected using the buffer gas cooling cleaning technique~\cite{1991_Savard}. An additional Ramsey cleaning \cite{2008_Eronen} step in the precision trap was needed to remove $^{135}$Xe$^+$, $^{135m}$Xe$^+$ and $^{135}$I$^+$. After the purification process a contamination level on the order of $1~\%$ of the data was observed in the measurement trap. However, the contaminant ions were well separated from the ions of interest by the PI-ICR technique~\cite{Nesterenko2018} and gated away for the analysis. A detailed ion rate dependency analysis~\cite{Kellerbauer2003} to probe for frequency shifts as a function of ion number did not show any significant deviations in the results.



Both the Time-of-Flight Ion-Cyclotron Resonance (ToF-ICR) \cite{Graff1980, Konig1995} technique utilizing Ramsey's method of time-separated oscillatory fields \cite{Kretzschmar_1999, George_2007_bis} and the newly commissioned PI-ICR \cite{Eliseev2014,Nesterenko2018} method were used for the mass-difference ($Q$-value) measurement. Both of the techniques provide the free-space cyclotron frequency
\begin{equation} \label{eq:qbm}
    \nu_c = \frac{1}{2\pi}\frac{q}{m}B,
\end{equation}
where $q/m$ is the charge-to-mass ratio of the ion and $B$ the magnetic field. The $Q$-value formula is:
\begin{equation}
    Q = m_p - m_d = (R - 1)(m_d - m_e),
    \label{eq:Q-value}
\end{equation}
where $m_p$ and $m_d$ are the masses of the parent ($^{135}$Cs$(7/2^+)$) and daughter ($^{135}$Ba$(3/2^+)$) atom, $R = \frac{\nu_d}{\nu_p}$ is their cyclotron frequency ratio for singly charged ions and $m_e$ is the electron rest mass. Since $(R-1)<10^{-5}$, the 0.3\,keV/c$^2$ uncertainty in the mass of $^{135}$Ba$(3/2^+)$ \cite{Wang_2017} is not a limitation for high-precision measurement. As the measured doublet has the same mass value $A$, mass-dependent errors become negligible \cite{Roux2013}. Contribution from the atomic electron binding energies is on the order of eV and thus can be neglected here.

\begin{figure}[h!]
    \includegraphics[width=\linewidth]{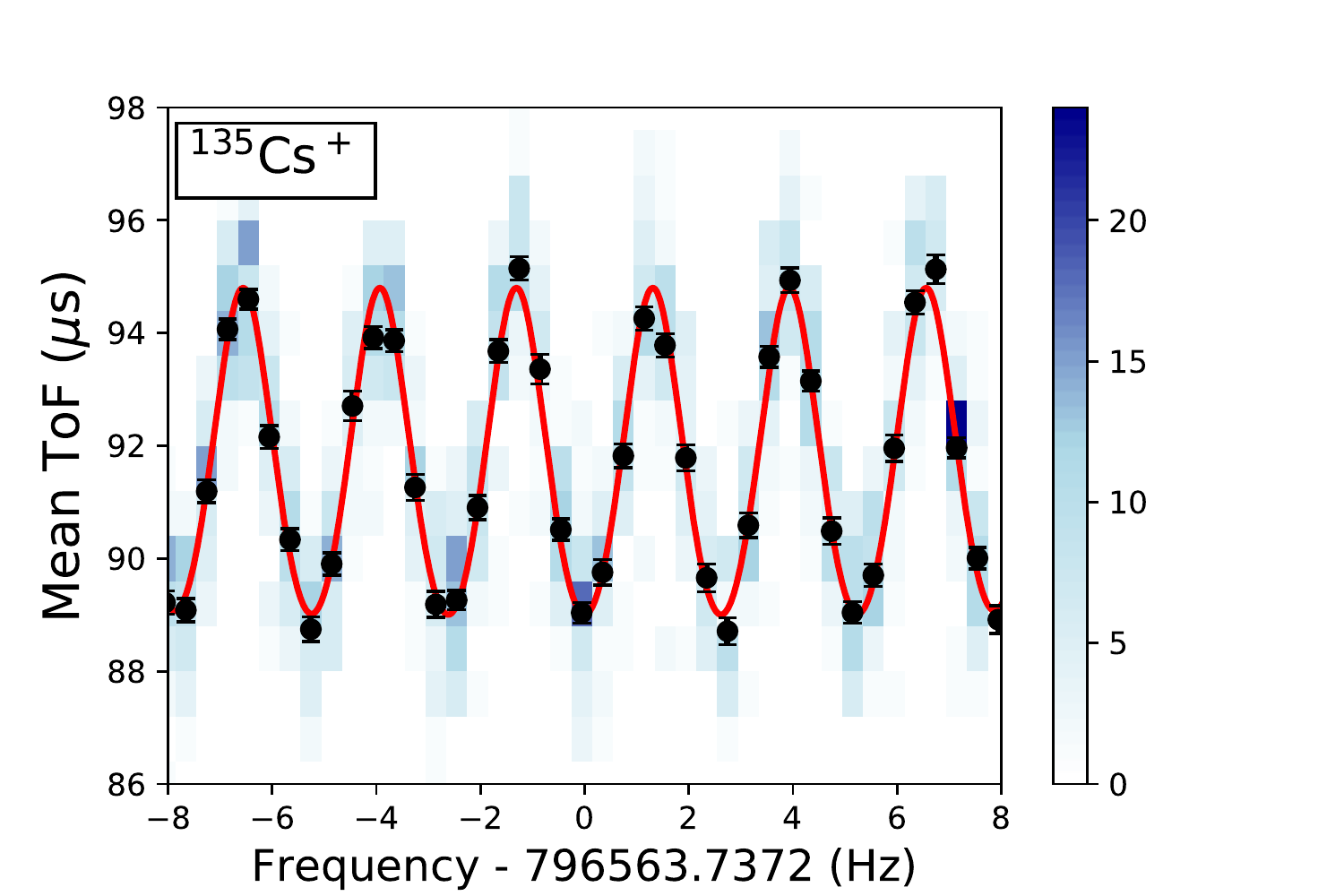}
    \caption{\label{fig:Ramsey_resonance}Ramsey ToF-ICR spectrum for $^{135}$Cs$^+$ ions using 25-350-25 ms (On-Off-On) excitation pattern. The mean data point are shown in black, the fit curve in red. The blue-shaded squares indicate the number of ions in each time-of-flight bin.}
\end{figure}

The Ramsey-type ToF-ICR cyclotron frequency measurements were performed for approximately 10~hours with a 25-350-25\,ms (On-Off-On) excitation pattern. The measurement was switched between parent and daughter ions every five scan rounds (about 2 minutes). A ToF-ICR resonance curve obtained using the Ramsey method is shown in Fig.~\ref{fig:Ramsey_resonance}.

\begin{figure}[h!]
    \centering
    \includegraphics[width=\linewidth]{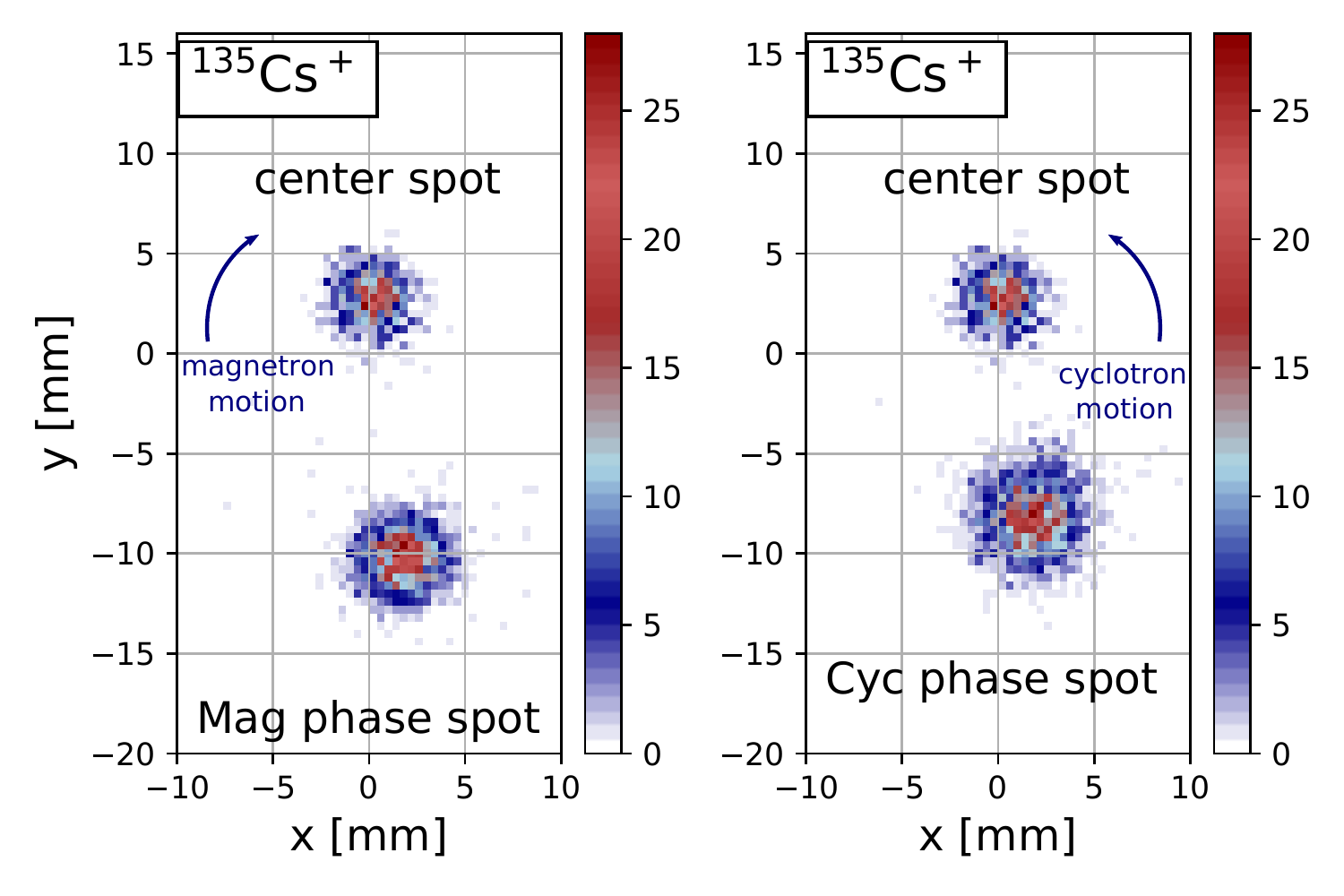}
    \caption{The three different ion-spots (center, magnetron phase and cyclotron phase) of $^{135}$Cs$^+$ on the 2D position-sensitive MCP detector after a typical PI-ICR excitation pattern. On the left is shown the magnetron phase spot, on the right the cyclotron phase spot. The angle difference between the two spots is related to the cyclotron frequency of the ion species. The number of ions in each pixel is indicated by color bars (colors in the online version).}
    \label{fig:cs135_4png}
\end{figure}

Data with the PI-ICR technique were collected for about 18 hours. The two phase spots, ``magnetron'' and ``cyclotron'', left and right panel in Fig.~\ref{fig:cs135_4png}, respectively, were collected using the timing patterns as described in \cite{Nesterenko2018}. The two phase spots were collected consecutively to account for any temporal shifts in the ion positions. The center spot is obtained by storing the ions in the trap for a few milliseconds and then extracting them. The extraction delay was varied over one magnetron period to account for any residual magnetron motion that could shift the different spots. The center spots were collected in approximately every three-hours intervals. The parent and daughter ion measurements were switched every few minutes. More information about the PI-ICR technique and the corresponding analysis can be found in \cite{Eliseev2014, Nesterenko2018}.

%
%

The ToF-ICR and PI-ICR data were split to 3 and 8 parts, respectively, for final fitting. Both types of measurements were checked for any count-rate related frequency shifts \cite{Kellerbauer2003}. Since no such shifts were observed, all bunches with up to 5 ions were used in the analysis. Temporal fluctuations of the $B$ field contribute less than $10^{-10}$ to the final frequency ratio uncertainty since the parent-daughter measurements were interleaved every few minutes \cite{Canete2016}. Likewise, frequency shifts in the PI-ICR measurement due to ion image distortions are well below the statistical uncertainty and thus were not added to the final uncertainty.





\begin{figure}[h!]
    \centering
    \includegraphics[width=\linewidth]{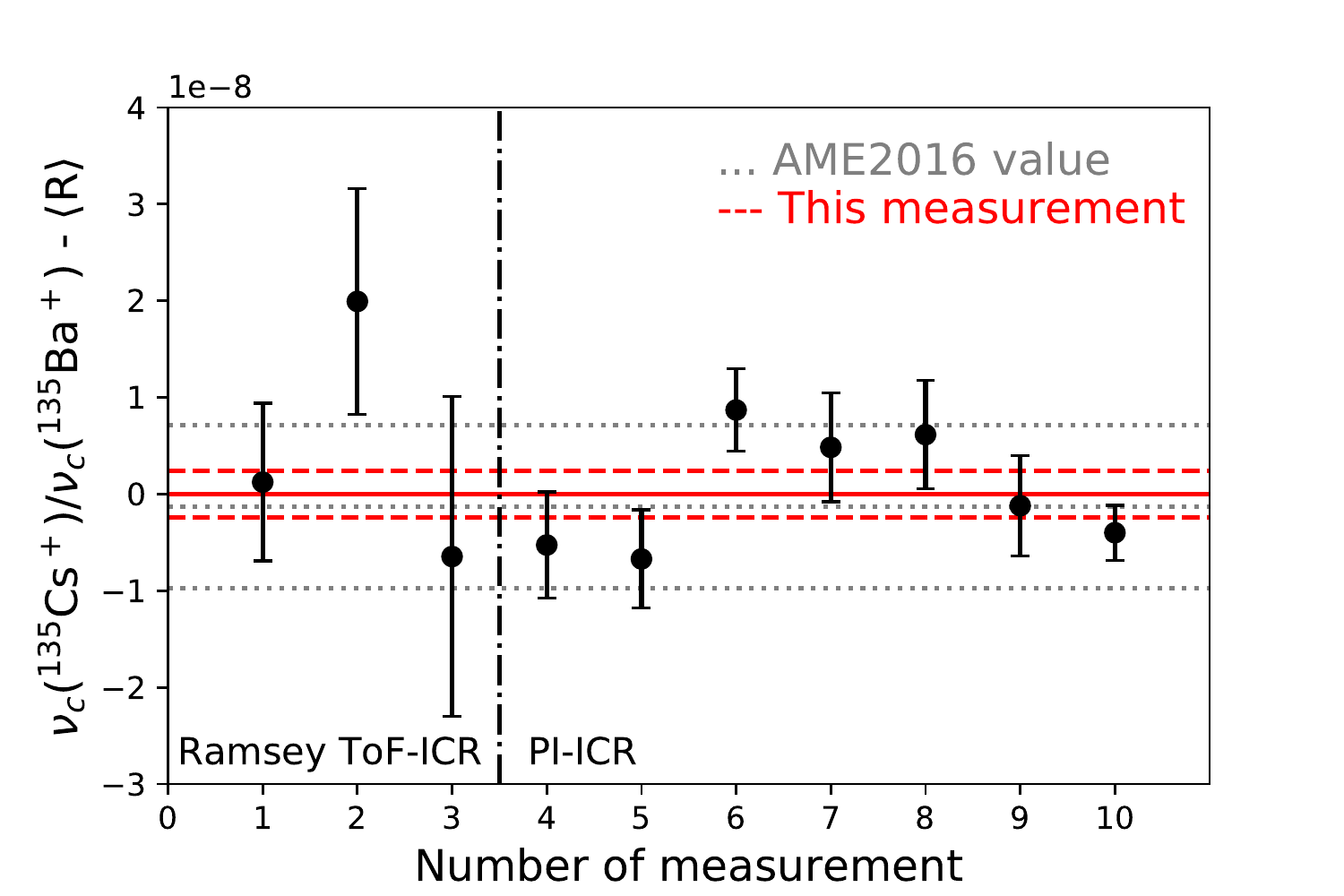}
    \caption{\label{fig:results}Difference between the cyclotron frequency ratios $\nu_c$($^{135}$Cs$^+$)/$\nu_c$($^{135}$Ba$^+$) measured in this work, shown as black data points, and the weighted average value from this work $\langle$R$\rangle = $ 1.0000021380(24) represented by the solid red line and its uncertainty in red dashed lines. The dotted gray lines represent the difference between our new value and the one referred to in the AME2016~\cite{Wang_2017} with its uncertainty. Colors in the online version.}
\end{figure}

The results of the analysis, including all data from both Ramsey-type ToF-ICR and PI-ICR measurements with comparison to literature values, are plotted in Fig.~\ref{fig:results}. The final results for the mean cyclotron frequency ratio between the daughter and parent nuclei and the corresponding $Q$-value are compared to literature values in Table~\ref{tab:results}.

\begin{table}[h!]
    \caption{\label{tab:results} Final result from the analysis with $\langle$R$\rangle$ being the mean cyclotron frequency ratio between the daughter and parent nuclei. The corresponding $Q$-value is also given as well as its comparison to the $Q$-value referred in the AME2016~\cite{Wang_2017}.}
    \centering
    \begin{tabular*}{\linewidth}{l@{\extracolsep{\fill}}l}
    \hline \hline \vspace{-0.25cm}
        \\
        $\langle$R$\rangle = \frac{\nu_{c,d}}{\nu_{c,p}}$ & 1.0000021380(24) \\ 
        $Q_{\beta^-}$ (this work)                    & 268.66(30)\,keV \\
        $Q_{\beta^-}$ (AME2016 \cite{Wang_2017})     & 268.9(10)\,keV \\
        $Q_{\beta^-}$ (\cite{Sugarman1949})          & 210(10)\,keV \\
        
    \hline \hline 
    \end{tabular*}
\end{table}

The new $Q$-value is a factor three more precise than that derived from masses of $^{135}$Cs and $^{135}$Ba given in AME2016 (\cite{Wang_2017} and references therein). In AME2016, the mass of $^{135}$Cs is tied to the very precisely known $^{133}$Cs mass through ($n$,$\gamma$) measurements and determines the mass of $^{135}$Cs with a weight of nearly 100\%. The mass of $^{135}$Ba is derived from ($n$,$\gamma$) links between $^{134}$Ba---$^{135}$Ba and $^{135}$Ba---$^{136}$Ba, which contribute to the $^{135}$Ba mass by 54.9\% and 45.1\%, respectively. Through Penning trap $Q$-value measurements,  $^{136}$Ba mass links to the mass of the $^{136}$Xe, which is known very precisely. $^{134}$Ba links to $^{133}$Cs through the $\beta$-decay of $^{134}$Cs and ($n$,$\gamma$). Additionally, we confirm that the value reported in \cite{Sugarman1949} is too low by 59(10) keV.




The GS-to-GS $Q$-value measured in this work is equal to 268.66(30)\,keV. This value allows for scenario I presented in \cite{Mustonen2011}, confirming that both the second-forbidden unique transition to the first excited state ($1/2^+$) and the first-forbidden unique transition to the second excited state ($11/2^-$) in $^{135}$Ba can occur with $Q$-values of 47.69(31)\,keV and 0.44(31)\,keV, respectively. 
Decay to the $3/2^+$ ground state of $^{135}$Ba has a half-life $(1.3-1.6)\times 10^6$ y \cite{MacDonald2016}. With the presently computed half-life estimate $(1-300)\times 10^{11}$ y (see below) for the transition to the $11/2^-$ state, the branching to this state is about $(0.04-16)\times 10^{-6}$. This branching ratio is close to that measured for the ultra-low-$Q$-value $\beta^-$ transition $^{115}\textrm{In}(9/2^+)\to\,^{115}\textrm{Sn}(3/2^+)$ in \cite{Wieslander2009}, $1.1 \times 10^{-6}$. Hence, it is feasible to detect the $^{135}\textrm{Cs}(7/2^+)\to\,^{135}\textrm{Ba}(11/2^-)$ transition. This and the fact that the transition has a simple unique (universal) shape of the electron spectrum and an ultra-low $Q$-value of only $0.44(31)$ keV make this transition an excellent candidate for neutrino-mass measurements.

\begin{figure}[h!]
 \begin{center}
    \includegraphics[width=\linewidth]{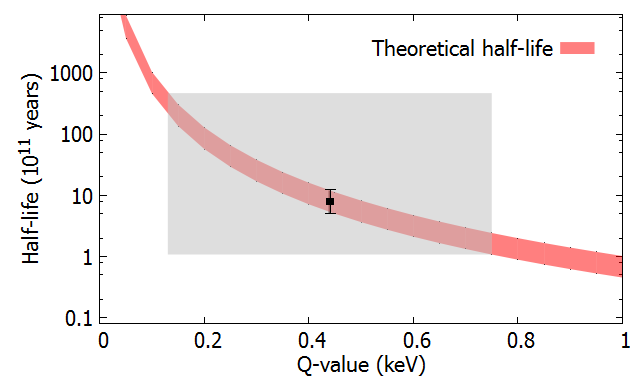}
    \caption{\label{fig:11hl} Shell-model calculated partial half-life of the 
decay of $^{135}$Cs to the second excited state in $^{135}$Ba as a function of 
the $Q$-value newly obtained: The red band corresponds to the range $g_{\rm A}^{\rm eff}=0.8-1.2$
of values of the axial coupling. The gray horizontal stripe gives the half-life 
assuming the best estimate $Q=0.44(31)$ keV from the present work. }
 \end{center}
\end{figure}

In order to estimate the partial half-lives for the transitions to the excited 
states, we have run large-scale shell-model calculations using the computer code
NuShellX@MSU \cite{nushellx}. The calculations were done in a model space 
consisting of the orbitals $0g_{7/2}$, $1d$, $2s$, and $0h_{11/2}$ for both protons 
and neutrons with the effective Hamiltonian Sn100pn \cite{sn100pn}. The shell-model 
calculations performed here represent a significant improvement over those 
performed in \cite{Mustonen2011}, where the microscopic quasiparticle-phonon
model (MQPM) \cite{mqpm1,mqpm2} was used to compute the involved nuclear wave
functions. Since the present shell-model calculations are much more sophisticated
than the old MQPM calculations the computational time required is several thousand 
times that needed in \cite{Mustonen2011}. The uncertainties related to the 
theoretical half-lives stem mostly from the unknown effective value, $g_{\rm A}^{\rm eff}$, 
of the axial-vector coupling. In Fig.~\ref{fig:11hl} we plot the partial half-life
of the transition to the $11/2^-$ state as a function of the $Q$-value.
The red band corresponds to the conservative
interval $g_{\rm A}^{\rm eff}=0.8-1.2$ deduced from a large body of related investigations
\cite{Suhonen2017}. Since the decays to the excited state are forbidden unique,
the half-lives are simply proportional to $g_{\rm A}^{-2}$, and one can easily
derive half-life estimates for any choice of $g_{\rm A}^{\rm eff}$.

\begin{table}[h!]
    \caption{\label{tab:hl} The $Q$-values and half-lives for the transitions of $^{135}\rm Cs(7/2^+)$ to the two excited states in $^{135}$Ba. The uncertainty for the transition to the $11/2^-$ state includes the nuclear structure only. The uncertainty related to the $Q$-value is one order of magnitude for this transition, while it is negligible for the transition to the $1/2^+$ state. }
    \centering
    \begin{tabular*}{\linewidth}{l@{\extracolsep{\fill}}ll}
    \hline \hline \vspace{-0.25cm}
        \\
		Transition to & $^{135}\rm Ba (1/2^+)$ & $^{135}\rm Ba (11/2^-)$ \\
		\hline
        $Q$-value (keV) & 47.69(31) & 0.44(31) \\
        $T_{1/2}$(ISM) (y)& $6.5(17) \times 10^{13}$ & $8.2(32) \times 10^{11}$ \\
        $T_{1/2}$(MQPM) (y) \cite{Mustonen2011} & $ 2 \times 10^{15}$ & $3\times 10^{10}$ \\
    \hline \hline
    \end{tabular*}
\end{table}

The calculated partial half-lives of the excited-state transitions are given 
in Table \ref{tab:hl}. Here the uncertainties arise from the assumed
interval $g_{\rm A}^{\rm eff}=0.8-1.2$ for the axial coupling.
Also the MQPM-computed half-lives, deduced from \cite{Mustonen2011}, are 
given for comparison. As can be seen, the presently computed partial half-lives
deviate substantially from those deduced from \cite{Mustonen2011}.
It should be noted that the ratio of the two half-lives, about 100 for the
present calculation and about five orders of magnitude for the calculation
of Mustonen \textit{et al.} \cite{Mustonen2011}, depends on two competing features.
The one unit of difference in the forbiddenness makes the decay to the
$11/2^-$ state some four orders of magnitude faster than the decay to the $1/2^+$
state \cite{Kostensalo2017}. On the other hand, the roughly 100 times larger
$Q$-value of the $1/2^+$ transition makes this transition faster by a couple 
orders of magnitude \cite{Behrens}, the net effect being that the transition to
the $11/2^-$ state can be estimated to be a rough two orders of magnitude
faster than the transition to the $1/2^+$ state, in agreement with the
results of the present shell-model calculation.

In conclusion, the $\beta$-decay $Q$-value of the transition from the $7/2^+$ ground
state of $^{135}$Cs to the $3/2^+$ ground state of $^{135}$Ba was measured 
with high precision at the JYFLTRAP Penning trap setup. This is the first direct determination of the $Q$-value. 
The new precise measurement confirms that the $Q$-value of the $\beta^-$-decay  transition $^{135}\textrm{Cs}(7/2^+)\to\,^{135}\textrm{Ba}(11/2^-)$ is positive with an ultra-low $Q$-value of $0.44(31)$ keV. Hence, this first-forbidden unique
transition, with a simple universal spectral shape, has the potential to serve as a 
candidate for antineutrino-mass measurements
with an almost order of magnitude lower $Q$-value than in 
presently running or planned direct (anti)neutrino-mass experiments.

This work has been supported by the Academy of Finland under the
Finnish Centre of Excellence Programme 2012-2017 (Nuclear and Accelerator Based Physics Research at JYFL) and projects No. 306980, 312544, 275389, 284516, 295207. This work was
supported by the EU Horizon 2020 research and innovation program under grant No. 771036 (ERC CoG MAIDEN).

\bibliographystyle{apsrev4-2}
\bibliography{Cs135_Qvalue}

\end{document}